\documentclass{ws-ijmpa}

\begin{document}

\markboth{P. Aschieri, E. Di Grezia, G. Esposito}
{Non-commutative Einstein equations and Seiberg--Witten map}

\catchline{}{}{}

\title{NON-COMMUTATIVE EINSTEIN EQUATIONS AND SEIBERG--WITTEN MAP}

\author{PAOLO ASCHIERI,$^{1,2}$ ELISABETTA DI GREZIA,$^{3}$
GIAMPIERO ESPOSITO$^{3}$}

\address{${ }^{1}$Dipartimento di Scienze e Tecnologie Avanzate,
Universit\`a del Piemonte Orientale, Italy\\
${ }^{2}$INFN, Sezione di Torino, Gruppo Collegato
di Alessandria, Italy\\
${ }^{3}$INFN, Sezione di Napoli,
Complesso Universitario di Monte S. Angelo,\\ Via Cintia, Edificio 6,
80126 Napoli, Italy\\
Received 15 June 2011}

\maketitle

\begin{abstract}
The Seiberg--Witten map is a powerful tool in non-commutative field
theory, and it has been recently obtained in the literature for
gravity itself, to first order in non-commutativity. This paper, relying
upon the pure-gravity form of the action functional considered in
Ref. 2, studies the expansion to first order of the
non-commutative Einstein equations, and whether the Seiberg--Witten map
can lead to a solution of such equations when the underlying classical
geometry is Schwarzschild.
\end{abstract}

\section{Introduction}

Non-commutative gravity theories are receiving much attention in the
literature because they are part of a promising research program aimed
at developing an algebraic approach to the longstanding problem of
quantum gravity.$^{1}$ In particular, we are here concerned with the
developments described in Refs. 2 and 3.

The work in Ref. 2 has built a geometric theory
of non-commutative gravity where the Lagrangian is a globally defined
4-form, invariant under diffeomorphisms as well as $*$-diffeomorphisms
and where in the commutative limit only the classical field 
degrees of freedom survive.
For pure gravity the action functional reads as
\begin{equation}
S=\int {\rm Tr}\Bigr(i \hat R \wedge_{*}\hat V 
\wedge_{*}\hat V \gamma_{5}\Bigr),
\label{(1)}
\end{equation}
where $\hat V$ is the tetrad 1-form and $\hat R$ is the curvature 2-form
\begin{equation}
\hat R=d \hat\Omega-\hat\Omega \wedge_{*}\hat \Omega.
\label{(3)}
\end{equation}
As in Ref. 2 we expand the tetrad on the basis of
$\gamma$-matrices,
\begin{equation}
\hat  V=\hat V^a\gamma_{a}+\widetilde {\hat V}{}^a\gamma_{a}\gamma_{5}.
\label{(2)}
\end{equation}
On denoting by $\eta_{ab}={\rm diag}(1,-1,-1,-1)$ the Minkowski
metric, and defining $\gamma_{ab} \equiv \gamma_{a}\gamma_{b}-I
\eta_{ab}$, one similarly expands the spin-connection 1-form
\begin{equation}
\hat\Omega={1\over 4}\hat\omega^{ab}\gamma_{ab}+i \hat\omega
I +{\widetilde {\hat\omega}}\gamma_{5}, 
\label{(4)}
\end{equation}
where $\hat \omega=\hat\omega_{\mu}dx^{\mu}$ and ${\widetilde {\hat\omega}}
={\widetilde {\hat \omega}}_{\nu}dx^{\nu}$ are 1-forms. From (3) and (4),
one has
\begin{equation}
\hat R={1\over 4}\hat R^{ab}\gamma_{ab}+i \hat r I +{\widetilde {\hat r}} \gamma_{5},
\label{(5)}
\end{equation}
where $\hat r$ and ${\widetilde {\hat r}}$ are 2-forms. We here correct some
numerical factors in Eqs. (5.2) and (5.3) of Ref. 2 and write the
following explicit formulae for the components of 
the curvature in Eq. (5),
\begin{eqnarray}
\hat R^{ab}&=& d \hat\omega^{ab}-{1\over 2} \hat\omega_{\; c}^{a} \wedge_{*}
\hat\omega^{cb}+{1\over 2}\hat \omega_{\; c}^{b} \wedge_{*}\hat \omega^{ca}
-i(\hat\omega^{ab} \wedge_{*}\hat\omega+\hat\omega \wedge_{*}\hat\omega^{ab})
\nonumber \\
&-& {i\over 2}\varepsilon_{\; \; \; cd}^{ab}
\Bigr(\hat\omega^{cd}\wedge_{*}{\widetilde {\hat \omega}}
+{\widetilde {\hat\omega}} \wedge_{*}\hat\omega^{cd}\Bigr),
\label{(6)}
\end{eqnarray}
\begin{equation}
\hat r=d \hat\omega-i \left({1\over 8}\hat\omega^{ab} 
\wedge_{*} \hat\omega_{ab}
+\hat\omega \wedge_{*} \hat\omega -{\widetilde {\hat\omega}} \wedge_{*}
{\widetilde {\hat\omega}} \right),
\label{(7)}
\end{equation}
\begin{equation}
{\widetilde {\hat r}}=d{\widetilde {\hat\omega}}
-i \Bigr(\hat\omega \wedge_{*}{\widetilde {\hat \omega}}
+{\widetilde {\hat\omega}}\wedge_{*}\hat\omega \Bigr)
+{i \over 16}\varepsilon_{abcd}\hat\omega^{ab} \wedge_{*}
\hat\omega^{cd}.
\label{(8)}
\end{equation}
We explicitly write in components the equations obtained by
varying the action (\ref{(1)}) with respect to 
the tetrad components $\hat V^a$
and ${\widetilde {\hat V}}{}^a$, they respectively read as
\begin{eqnarray} &\;&
-\Bigr(\hat V^{d}\wedge_{*}\hat R^{ab}+\hat R^{ab}\wedge_{*}\hat V^{d}\Bigr)
\varepsilon_{abcd}+i(\eta_{bc}\eta_{ad}-\eta_{ac}\eta_{bd})
\Bigr(\hat R^{ab}\wedge_{*}{\widetilde {\hat V}}{}^{d}
-{\widetilde {\hat V}}{}^{d}
\wedge_{*}\hat R^{ab}\Bigr) \nonumber \\
& &~~+ 4i \eta_{dc}\Bigr({\widetilde {\hat r}}\wedge_{*}\hat V^{d}
-\hat V^{d}\wedge_{*}{\widetilde {\hat r}}\Bigr)
+4 \eta_{dc}\Bigr({\widetilde {\hat V}}{}^{d}\wedge_{*}{\hat r}
+{\hat r} \wedge_{*}{\widetilde {\hat V}}{}^{d}\Bigr)=0,
\label{(9)}
\end{eqnarray}
\begin{eqnarray}
&\;& -\Bigr({\widetilde {\hat V}}{}^{d}\wedge_{*}\hat R^{ab}
+\hat R^{ab}\wedge_{*}{\widetilde {\hat V}}{}^{d}\Bigr)
\varepsilon_{abcd}+i(\eta_{bc}\eta_{ad}-\eta_{ac}\eta_{bd})
\Bigr(\hat R^{ab}\wedge_{*}\hat V^{d}-\hat V^{d}
\wedge_{*}\hat R^{ab}\Bigr) \nonumber \\
& &~~+ 4i \eta_{dc}\Bigr({\widetilde {\hat r}}\wedge_{*}
{\widetilde {\hat V}}{}^{d}
-{\widetilde {\hat V}}{}^{d}\wedge_{*}{\widetilde {\hat r}}\Bigr)
+4 \eta_{dc}\Bigr(\hat V^{d}\wedge_{*} \hat r
+\hat r \wedge_{*}\hat V^{d}\Bigr)=0.
\label{(10)}
\end{eqnarray}
We notice that Eq. (10) can be obtained by the replacements $\hat V^{a}
\rightarrow {\widetilde {\hat V}}{}^{a}$ and ${\widetilde {\hat V}}{}^{b}
\rightarrow \hat V^{b}$ in Eq. (9).

The torsion 2-form is defined by
\begin{equation}
\hat T \equiv d\hat V-\Omega \wedge_{*}\hat V -\hat V \wedge_{*}\hat \Omega,
\label{(11)}
\end{equation}
and expanding $\hat T$ on the basis of $\gamma$-matrices as in
Eq. (2) we have $\hat T=\hat T^{a}\gamma_{a}
+{\widetilde {\hat T}}{}^{a}\gamma_{a}\gamma_{5}$.
Variation of the action with respect to the spin-connection gives
the remaining field equation $\left \{ \hat T \wedge_{*}\hat V-\hat V
\wedge_{*}\hat T,\gamma_{5} \right \}=0$, where we use curly brackets
for anticommutators. Since $\hat T \wedge_{*}\hat V$ is even in the
$\gamma$-matrices we eventually obtain
\begin{equation}
\hat T \wedge_{*}\hat V-\hat V \wedge_{*}\hat T=0, \label{(12)}
\end{equation}
which is satisfied by a vanishing torsion.

{\vskip.4cm}
The work in Ref. 3 has studied the
Seiberg--Witten map$^{4}$ which relates non-commutative
degrees of freedom for spin-connection, tetrad and gauge
parameter to their commutative counterparts. The relation is such that
non-commutative gauge transformations
(with non-commutative gauge parameter $\hat\Lambda$) correspond to
commutative gauge transformations (with commutative gauge parameter
$\Lambda$). For example for the tetrad we have
\begin{equation}
{\widehat V}_{\mu}+\delta_{\widehat \Lambda}{\widehat V}_{\mu}
={\widehat V}_{\mu}(V+\delta_{\Lambda}V,
\omega+\delta_{\Lambda}\omega),
\label{(17)}
\end{equation}
where $\hat V=\hat V_\mu dx^\mu$ is the non-commutative tetrad 
1-form, while  $V=V^a\gamma_a=V^a_\mu dx^\mu\gamma_a$ and
$\omega={1\over 4}\omega^{ab}\gamma_{ab}= {1\over
4}\omega_\mu^{ab}dx^\mu\gamma_{ab}$ are the usual commutative
tetrad and spin-connection. The non-commutativity we consider  
is given by the Moyal--Weyl
$\star$-product associated with a constant antisymmetric matrix
$\theta^{\rho\sigma}$ in the (not necessarily Cartesian) 
coordinates $x^\mu$; we have 
${x}^{\rho}\star{ x}^{\sigma}-{x}^{\sigma}\star{ x}^{\rho}=i \theta^{\rho
 \sigma}$.
Equation (\ref{(17)}) can be solved
order by order in perturbation theory. The solution of Eq. (\ref{(17)})
to first order in the non-commutativity $\theta^{\rho \sigma}$ has the
general structure
\begin{equation}
{\widehat V}_{\mu}={\widehat V}_{\mu}^{(0)a} \gamma_{a}
+{\widehat V}_{\mu 5}^{(1)a}\gamma_{5}\gamma_{a}
+{\rm O}(\theta^{2}).
\label{(18)}
\end{equation}
The zeroth- and first-order terms in $\theta^{\rho\sigma}$ 
in Eq. (\ref{(18)}) are found to be$^{3}$
\begin{equation}
{\widehat V}_{\mu}^{(0)a}=V_{\mu}^{a},
\label{(19)}
\end{equation}
\begin{equation}
{\widehat V}_{\mu 5}^{(1)a}\equiv-{\widetilde V}_{\mu}^{a}
={1\over 4}\theta^{\lambda \sigma}\omega_{\lambda}^{eb}
\left(\partial_{\sigma}V_{\mu}^{c}-{1\over 2}
\omega_{\sigma}^{cd}V_{\mu d} \right)
\varepsilon_{ebc}^{\; \; \; \; \; a}.
\label{(20)}
\end{equation}
Similarly, on writing for the spin-connection
\begin{equation}
{\widehat \omega}_{\mu}={1\over 2}
{\widehat \omega}_{\mu}^{(0)ab}\sigma_{ab}
+{\widehat a}_{\mu}^{(1)}
+i{\widehat b}_{\mu 5}^{(1)}\gamma_{5},
\label{(21)}
\end{equation}
one finds,$^{3}$ to first order in $\theta^{\rho \sigma}$,
\begin{equation}
{\widehat \omega}_{\mu}^{(0)ab}=\omega_{\mu}^{ab},
\label{(22)}
\end{equation}
\begin{equation}
{\widehat a}_{\mu}^{(1)}\equiv -i \omega_{\mu}
=-{1\over 16}\theta^{\lambda \sigma}\omega_{\lambda}^{ab}
\Bigr(\partial_{\sigma}\omega_{\mu}^{cd}
+R_{\sigma \mu}^{cd}\Bigr)
\eta_{ac}\eta_{bd},
\label{(23)}
\end{equation}
\begin{equation}
{\widehat b}_{\mu 5}^{(1)} \equiv i {\widetilde \omega}_{\mu}
=-{1\over 32}\theta^{\lambda \sigma}\omega_{\lambda}^{ab}
\Bigr(\partial_{\sigma}\omega_{\mu}^{cd}
+R_{\sigma \mu}^{cd}\Bigr)
\varepsilon_{abcd}.
\label{(24)}
\end{equation}

In this note we consider the non-commutative fields appearing in the
action (1) and in the corresponding field equations (9) and (10) as
dependent on the commutative ones via Seiberg--Witten map. Hence 
we expand equations (9) and (10) in terms of the commutative tetrad and
spin-connection. This is done to first order in $\theta^{\mu \nu}$ by inserting
the first-order Seiberg--Witten map (15), (16) and (18)--(20). We have
done so when the underlying classical geometry is the spherically
symmetric solution of the classical Einstein equations in vacuum, i.e.
Schwarzschild. 

\section{Expansion to first order of the non-commutative Einstein
equations}

By virtue of the wedge-$*$ product of forms defined in Ref. 2,
for any 1-form $\alpha^{1},\beta^{1}$ and any
2-form $\gamma^{2}$, one can write (provided that 
$\partial_\mu x^\rho=\delta_\mu^\rho$),
\begin{equation}
\alpha^{1} \wedge_{*}\beta^{1}=\alpha^{1} \wedge \beta^{1}
+{i\over 2}\theta^{\rho \sigma}
\Bigr(\partial_{\rho}\alpha_{\mu}^{1}\Bigr)
\Bigr(\partial_{\sigma}\beta_{\nu}^{1}\Bigr)
dx^{\mu} \wedge dx^{\nu}+{\rm O}(\theta^{2}),
\label{(25)}
\end{equation}
\begin{equation}
\alpha^{1} \wedge_{*} \gamma^{2} = \alpha^{1} \wedge \gamma^{2}
+{i\over 2}\theta^{\rho \sigma}\Bigr(\partial_{\rho}\alpha_{\mu}^{1}\Bigr)
\Bigr(\partial_{\sigma}\gamma_{\nu \lambda}^{2}\Bigr)
dx^{\mu} \wedge dx^{\nu} \wedge dx^{\lambda}
+{\rm O}(\theta^{2}).
\label{(26)}
\end{equation}
Eq. (\ref{(26)}) can be applied repeatedly to the $\theta$-expansion  of Eqs.
(9) and (10). For this purpose we need from Eq. (\ref{(26)}) the identities
\begin{equation}
\alpha^{1} \wedge_{*}\gamma^{2}+\gamma^{2}\wedge_{*}\alpha^{1}
=2 \alpha^{1} \wedge \gamma^{2}+{\rm O}(\theta^{2}),
\label{(27)}
\end{equation}
\begin{equation}
\alpha^{1}\wedge_{*}\gamma^{2}-\gamma^{2}\wedge_{*}\alpha^{1}
=i \theta^{\rho \sigma}\Bigr(\partial_{\rho}\alpha_{\mu}^{1}\Bigr)
\Bigr(\partial_{\sigma}\gamma_{\nu \lambda}^{2}\Bigr)
dx^{\mu} \wedge dx^{\nu} \wedge dx^{\lambda}+{\rm O}(\theta^{2}).
\label{(28)}
\end{equation}
Moreover, from the work in Ref. 2 we know that in non-commutative
field theory charge-conjugation conditions imply that
\begin{equation}
{\widetilde {\hat V}}{}^{a}(\theta)=-{\widetilde {\hat V}}{}^{a}(-\theta), \;
\hat \omega(\theta)=-\hat \omega(-\theta), \;
{\widetilde {\hat \omega}}(\theta)=-{\widetilde {\hat \omega}}(-\theta),
\label{(29)}
\end{equation}
and hence all non-commutative fields that are not present in the
commutative case are at least proportional to $\theta$ (and hence
vanish in the commutative limit),
\begin{equation}
{\widetilde{\hat V}}_{\mu}{}^{a}={\rm O}(\theta), \;
\omega={\rm O}(\theta), \;
{\widetilde {\hat \omega}}={\rm O}(\theta), \;
\hat  r={\rm O}(\theta), \;
{\widetilde {\hat r}}={\rm O}(\theta).
\label{(30)}
\end{equation}
By virtue of (\ref{(27)}), (\ref{(28)}) and (\ref{(30)}), 
Eq. (\ref{(9)}) reduces to
\begin{equation}
\varepsilon_{abcd}V^{d} \wedge \Bigr(R^{(0)ab}+R^{(1)ab}\Bigr)
+{\rm O}(\theta^{2})=0,
\label{(31)}
\end{equation}
where $R^{(n)ab}$ denotes the $n$-th order part of the curvature
2-form in powers of $\theta^{\rho \sigma}$, while Eq. (\ref{(10)}) becomes
\begin{equation}
\Bigr[-\varepsilon_{abcd}{\widetilde V}_{\mu}^{d}R_{\nu \lambda}^{(0)ab}
+\theta^{\rho \sigma}\Bigr(\partial_{\rho}V_{\mu}^{d}\Bigr)
\Bigr(\partial_{\sigma}R_{dc \; \nu \lambda}^{(0)}\Bigr)
+4V_{c \mu}r_{\nu \lambda}^{(1)}\Bigr]
dx^{\mu} \wedge dx^{\nu} \wedge dx^{\lambda}
+{\rm O}(\theta^{2})=0,
\label{(32)}
\end{equation}
having defined
\begin{equation}
r_{\nu \lambda}^{(1)} \equiv \theta^{\rho \sigma}
r_{\nu \lambda \rho \sigma}^{(1)}.
\label{(33)}
\end{equation}
We note that, in Eq. (\ref{(31)}), since $V^{d}$ is the classical tetrad,
the term $\varepsilon_{abcd}V^{d} \wedge R^{(0)ab}$ vanishes for
any solution of the vacuum Einstein equations.

\section{Non-commutative Einstein equations under the first-order
Seiberg--Witten map}

Now we consider the coordinates $x^1=t, x^2=r, x^3=\vartheta, x^4=\varphi$
and the tetrad $V^{a}=V_{\mu}^{a}dx^{\mu}$ given by
\begin{equation}
V^{(t)}=\sqrt{1-{2M \over r}}dt, \;
V^{(r)}={dr \over \sqrt{1-{2M\over r}}}, \;
V^{(\vartheta)}=r d\vartheta, \;
V^{(\varphi)}=r \sin \vartheta d\varphi.
\label{(34)}
\end{equation}
This tetrad with the associated spin-connection 
(see formula (34) below)
describe a  Schwarzschild geometry solution of the classical 
Einstein equations in first-order formalism.
{}For this tetrad we have  $R_{\mu \nu}^{(1)ab}=0$, 
so that Eq. (\ref{(31)}) becomes an
identity. In order to try to solve the first-order non-commutative
equations (\ref{(32)}) we consider for the first-order expression of the
non-commutative tetrad the expression given by Eq. (\ref{(20)}),
which is the first-order Seiberg--Witten map for the tetrad.
This ansatz is supported by the following argument:

{To first order in $\theta$ the non-commutative Einstein action (1) is
the same as the commutative one if the non-comutative fields are
re-expressed in terms of the commutative ones
by using the Seiberg--Witten map (indeed the non-commutative fields
satisfy the charge conjugation conditions of Ref. 2).
Therefore it is expected that the non-commutative Einstein equations
are automatically satisfied if the non-commutative fields are expressed
via Seiberg--Witten map in terms of the commutative ones.}

We then substitute Eq. (\ref{(20)})  into Eq. (\ref{(32)}), we relabel
the indices summed over, and assuming (for ease of
calculations) that only the
non-commutativity component  $\theta^{23}$
(that we rename $\theta$) is non-vanishing, i.e.
\begin{equation}
\theta^{\gamma \sigma} \frac{\partial}{\partial x^\gamma}\wedge
\frac{\partial}{\partial x^\sigma}= \theta
\biggr(\frac{\partial}{\partial r}\otimes \frac{\partial}{\partial
\vartheta} -\frac{\partial}{\partial \vartheta}\otimes
\frac{\partial} {\partial  r}\biggr), \label{(35)}
\end{equation}
we re-express Eq. (\ref{(32)}) in the form (with our notation $dx^1=dt,
dx^2=dr, dx^3=d\vartheta, dx^4=d\varphi$)
\begin{eqnarray}
\; & \; & \biggr \{
\varepsilon_{abcd}\varepsilon_{epq}^{\; \; \; \; \; d}
R_{\nu \lambda}^{(0)ab}\biggr[\Bigr(\partial_{3}V_{\mu}^{q}\Bigr)
\omega_{2}^{ep}-\Bigr(\partial_{2}V_{\mu}^{q}\Bigr)\omega_{3}^{ep}
+{1\over 2}V_{\mu f}\Bigr(\omega_{2}^{qf}\omega_{3}^{ep}
-\omega_{3}^{qf}\omega_{2}^{ep}\Bigr)\biggr] \nonumber \\
&~&~+ 4 \biggr[\Bigr(\partial_{2}V_{\mu}^{q}\Bigr)
\Bigr(\partial_{3}R_{qc \; \nu \lambda}^{(0)}\Bigr)
-\Bigr(\partial_{3}V_{\mu}^{q}\Bigr)
\Bigr(\partial_{2}R_{qc \; \nu \lambda}^{(0)}\Bigr)\biggr] \nonumber \\
&~&~+ 16V_{c \mu}\Bigr(r_{\nu \lambda 23}^{(1)}-r_{\nu \lambda 32}^{(1)}\Bigr)
\biggr \}dx^{\mu}\wedge dx^{\nu}\wedge dx^{\lambda}=0.
\label{(36)}
\end{eqnarray}
At this stage, the identity
\begin{equation}
\varepsilon^{abcp}\varepsilon_{defp}=-{\rm det}
\pmatrix{\delta_{d}^{a} & \delta_{d}^{b} & \delta_{d}^{c} \cr
\delta_{e}^{a} & \delta_{e}^{b} & \delta_{e}^{c} \cr
\delta_{f}^{a} & \delta_{f}^{b} & \delta_{f}^{c} \cr}
\label{(37)}
\end{equation}
can be exploited, jointly with the standard evaluation of classical
curvature 2-form and classical spin-connection 1-form, the latter
being given by
\begin{equation}
\omega_{\mu}^{ab}={1\over 2}V^{a \nu}\Bigr(V_{\nu,\mu}^{b}
-V_{\mu,\nu}^{b}\Bigr)-{1\over 2}V^{b\nu}
\Bigr(V_{\nu,\mu}^{a}-V_{\mu,\nu}^{a}\Bigr)
+{1\over 2}V^{a\nu}V^{b \sigma}\Bigr(V_{\nu,\sigma}^{c}
-V_{\sigma,\nu}^{c}\Bigr)V_{c \mu}.
\label{(38)}
\end{equation}
Hence we find in a Schwarzschild background, bearing also in mind
Eq. (\ref{(7)}), the Seiberg--Witten map for the spin-connection
Eq. (\ref{(23)}), and the definiton (\ref{(33)}), 
that the left-hand side of Eq. (\ref{(36)}) takes the form
$$
K_{c123}dx^{1}\wedge dx^{2} \wedge dx^{3}
+K_{c124}dx^{1}\wedge dx^{2} \wedge dx^{4}
+K_{c134}dx^{1}\wedge dx^{3} \wedge dx^{4}
+K_{c234}dx^{2}\wedge dx^{3} \wedge dx^{4},
$$
where each $K_{c \mu \nu \lambda}$ can be written as the sum of
$6$ terms. We obtain, after a lengthy calculation,
the simple formulae
\begin{equation}
K_{c123}={4M(5M-2r)\over r^{4}\sqrt{1-{2M \over r}}}\delta_{c1},
\label{(39)}
\end{equation}
\begin{equation}
K_{c124}=K_{c134}=0, \; \forall c=1,2,3,4,
\label{(40)}
\end{equation}
\begin{equation}
K_{c234}=4M {\sin \vartheta \over r^{2}}\delta_{c4}.
\label{(41)}
\end{equation}
Since $K_{1123}$ and $K_{4234}$ are non-vanishing, the
field configurations given in Eq. (\ref{(20)}), (\ref{(23)}) and
obtained by applying the Seiberg--Witten
map to the classical tetrad of Eq. (\ref{(34)}) and spin-connection of
Eq. (\ref{(38)}),  are not solutions of the
non-commutative Einstein equations.

In order to search for solutions to non-commutative Einstein
equations that in the
commutative limit become Schwarzschild we have therefore to revert to
Eq. (\ref{(32)}), where no use of the Seiberg--Witten map is made, and  look
for solutions of Eq. (\ref{(32)}) and also of the torsion
constraints in Eq. (\ref{(12)}).
\vskip.4cm
In conclusion, the performed calculation shows that there
is a mismatch between: (I) using the Seiberg--Witten map in the
non-commutative action (1) in order to express all non-commutative
fields in terms of the commutative tetrad and spin-connection, and
then solving the action (that in general will be a higher derivative
action) by varying with
respect to only the classical fields.\\
(II) obtaining the
non-commutative field equations by varying the action (1) with respect
to all non-commutative fields, and then trying to solve these equations by
expressing the non-commutative fields in terms of the commutative
ones via Seiberg--Witten map.

This mismatch can be due to the fact that in case (II), in order to obtain
the equations of motion we have to vary also with respect to the extra
fields ${\widetilde {\hat V}}$, ${\widetilde {\hat \omega}}$ and $\hat
\omega$.
Exactly these corresponding extra equations of motion are not
satisfied by considering the field configurations
${\widetilde {\hat V}}$, ${\widetilde {\hat \omega}}$ and $\hat \omega$
obtained by the Seiberg--Witten map with
$V^a$ and $\omega^{ab}$ the classical
black hole tetrad and spin-connection.
\vskip.4cm

We also notice that we have chosen the non-commutativity directions
$\frac{\partial}{\partial r}$ and $\frac{\partial}{\partial \vartheta}$
not to be Killing vector fields for our
classical black hole solution. This was done on purpose because,
similarly to Refs.$^{[5,6,7]}$, it is possible to show that,
when non-commutativity is (in part) obtained
by using Killing vector fields of a given classical solution to
Einstein equations, then this classical solution is also
a solution of the non-commutative field equations, where all
the extra non-commutative fields are taken to vanish.

\section*{Acknowledgments}

G. Esposito is grateful to the Dipartimento di Scienze Fisiche of
Federico II University, Naples, for hospitality and support; he dedicates
to Maria Gabriella his contribution to this work.
We are grateful to P. Vitale for scientific conversations.

\end{document}